# Adjusting the nuclear reactor's neutron transport and diffusion theory for an alternative description and modelling of postage or supplies delivery processes

Petropoulos N.P. [a]

Nuclear Engineering Laboratory, School of Mechanical Engineering, National Technical University of Athens, 15780 Athens, GREECE

## Abstract

There seems to exist significant similarities between a reactor system and a supply chain from collection to delivery. In the reactor case, neutrons are continuously produced and absorbed in nuclear fuel. In a supply system case, items are continuously collected and continuously delivered to destinations. Stable reactor operation is ensured by keeping the ratio of neutrons produced to neutrons absorbed in the reactor equal to one. Profitable and qualitative supply operation is ensured by keeping the ratio of items delivered to items collected as close to unity as possible. The analogy between the two systems is obvious. This text, which is provided as is and has not undergone any peer review process, proposes transferring parts of the nuclear reactor's neutron transport and diffusion theory to deterministically model supply processes. To this end a set of assumptions and definitions are provided as needed along with the introduction of reactions or interactions like collections, deliveries, and losses occurring into the supply chain. The interaction rates are calculated with the method used in reactors employing analogy factors and interactors with which the items in the chain interact. The main aim is to describe losses escape in steady state and in parallel estimate the analogy factors and optimize for the correct selection of the interactors pool. The model as proposed seems to be a tool for a different insight method into supply problems. The model, if proven and applied, is discussed to be a strong optimization tool, which could deterministically pinpoint flaws in existing supply systems or stochastically efficiently organize proposed supplied chains.

---

[a] npetr@mail.ntua.gr

## 1. Introduction

This study could be considered as a spin-off extension of Nuclear Engineering and Operational Reactor Physics into today's delivery enterprises. The engineering and atomic science masters of the 2$^{nd}$ third of the 20$^{th}$ century were utterly successful to describe systems as complex as the nuclear reactors using theory and models, which, surprisingly enough, have proven very simple, very effective and widely comprehensive. This achievement was accomplished when the present computing capabilities were absolutely unthinkable and, what is most important, when a rather significant part of the physical properties of nuclear fuel and neutron behavior in reactor matter were unknown. However, their approach overcame all information gaps and successfully filled the blanks to build reactors, which, under the same basic principles are the ones efficiently operating today.

It seems that there are significant similarities between a reactor system and a supply chain from collection to delivery. In the reactor case, a cloud of neutrons is continuously produced by fissions and through intended processes continuously absorbed in nuclear fuel within a confined space. Safe reactor operation is ensured by keeping the ratio $k_n$ of neutrons produced from fission to neutrons absorbed in the reactor equal to one. In a supply system case, a cloud of deliverable items is continuously produced from collections and through intended processes continuously delivered to destinations within a business volume, which for some enterprises is global, while for other, smaller ones, a confined balloon. Profitable business operation at a certified quality is ensured by keeping the ratio $k_s$ of items delivered to items collected or entered in the supply chain as close to unity as possible. The analogy between the two systems is more than obvious. Fortunately, unlike a nuclear reactor, a supply system cannot go supercritical (i.e. $k_s > 1$), thus making the analogy much simpler to consider.

In both cases, operations depend on time, yet since a supply business is usually operating at capacity, it can be accepted that modelling should focus to steady state

$$\frac{dk_s(t)}{dt} = 0$$

and that would be similar to the critical nuclear reactor. There are situations where

$$\frac{dk_s(t)}{dt} < 0$$

but, till now and on the long businesses run, these are exceptional, related to absolutely unexpected events, like the COVD-19 pandemic, volcano eruptions, major airport strikes, war states, customs bottle necks, border confrontations etc. Furthermore, capacity increases are technical events of non-interfering character, organized and promoted outside the supply chain operation. Once introduced (usually abruptly) into the chain, the increase of collected items would temporarily lead $k_s$ to low values < 1 until balance is reestablished. However, all these time related events are transient phenomena, which need another kind of modelling, if any at all.

In the nuclear reactor case, space considerations are important, when there is a need for detailed calculations regarding the performance of a nuclear reactor core. However, macroscopic or mesoscopic modelling, where space and dimensions play a secondary role, has been found to provide approximate results, which are not far from reality. It is envisaged that a similar approach could be used for supply chains. Some exceptions do exist and should be taken into account, if necessary.

Despite all mentioned similarities, adjusting the nuclear reactor's neutron transport and diffusion theory for an alternative description and modelling of postage or supplies delivery processes is certainly challenging. Initially, in this study, main assumptions and definitions are provided at the liberty of the author and according to needs. More assumptions and definitions are provided progressively. Definitions may not always be of the type usually involved in supply chain logistics, however, the model suggested originates from a totally different discipline and some room for coarser and even somewhat inaccurate interpretations should, in principle, be allowed. The new supply chain properties introduced, that is the so-christened interaction significance macro-factors are linear analogy coefficients, which, multiplied with items flow in the supply chain provide the respective interaction rates. Interactions are supposed to be promoted by respective interactors. Main interactions are, naturally, defined as collections, deliveries, and losses along the supply chain. The aim is to describe losses escape in steady state and thus calculate the correction factor, which multiplied with $k_s$ would make it unity. Collections are being considered as Poisson processes, yet some, hopefully new discussion and insight is introduced regarding the collection process and the first mile in general. A similar discussion is introduced for the last mile. Regarding losses escape, this, according to the present approach, is not defined for first mile processes but it does run during items forwarding to destination (i.e. middle mile) and delivery. Within middle mile, losses escape accounting is carefully analyzed and explained in a deterministic rather than stochastic manner. It is deduced that during middle mile, losses escape is certainly a percentage but its calculation goes far beyond a simple ratio. This called for a discussion of the forwarding properties of a supply system and the introduction of the entry and forwarding interaction macro-factor. Finally, and after a closer look for the last mile phase characteristics and properties arising from the viewpoint of Nuclear Engineering, losses escape at the last mile are predicted more simply as items delivered over items in the final distribution phase.

## 2. Main assumptions and definitions

In order to explain efficiently the proposed model, certain assumptions and definitions are needed. First of all, the supply process could be divided in three main and fairly discrete phases: (a) first mile, (b) middle mile: entry and forwarding, and (c) last mile, where all types of miles are well-accepted customary supply chain descriptions of the movement of goods from the customer to the collection post, in-between collection and distribution and from the distribution post to the final destination, respectively.

Secondly, there is a need to attribute a property or even properties to the items into the supply chain or into the transportation process, which could serve as independent variables for the proposed theory and modelling. A single such variable should be enough for a coarse discussion and, for the purposes of this study, it could be called "goods enthalpy". It is suggested that this particular enthalpy may be defined in three segments

(i) the first mile phase enthalpy "$H_a$":

$$H_a = \text{item's distance between customer's address to collection post} \times \text{item's weight} \times \text{item's volume} \times ... \times ... \times ... \quad (1)$$

(ii) the middle mile enthalpy "$H_b$", which is connected to all processes after the item's pickup or drop – off till its arrival to the distribution post:

$$H_b = \text{item's distance between collection post to distribution post} \times \text{item's weight} \times \text{item's volume} \times ... \times ... \times ... \quad (2)$$

and

(iii) the last mile enthalpy "$H_c$", which is connected to all processes after the item's arrival at the distribution post till its delivery to the client's address:

*$H_c$ = item's distance between distribution post to client's address × item's weight × item's volume × ... × ...× ...* (3)

In all cases, enthalpy would serve as a variable that describes the difficulties in the supply process. The higher the enthalpy, the more difficult and costly to accept the item in the chain and forward it to its destination. On the other hand, of course, if enthalpy is too low, item's processing into a supply chain might not worth it. Causal relations between enthalpy segments are certainly difficult to define or adjust when needed. In this study, any discussion over any possible integration of enthalpy $H_a$ with enthalpy $H_b$ is in principle not encouraged. Relevant reasoning is given in section 3. On the contrary, there is room for considering $H_c$ as an $H_b$ extension. For example, according to the reasonable hypothesis that [*item's distance between distribution post to client's address*] is far less than [*item's distance between collection post to distribution post*], one may replace enthalpy $H_b$ in Eq. 2 with enthalpy $H = H_b + H_c$. Therefore, Eq. 2 may be rewritten as:

*$H$ = item's distance between collection post to client's address × item's weight × item's volume × ... × ...× ...* (4)

More reasoning towards a strong connection between $H_c$ and $H_b$ is given in section 5.

Thirdly, one should describe with some acceptable efficiency, the item's interactions with and within the supply chain. These interactions are certainly more than a handful, however let us identify those, which seem more important, i.e.: (i) "*e*" chain entry, (ii) "*f*" forwarding (that is middle mile transport processes towards destination), (iii) "*d*" delivery and, of course and unavoidably, (iv) "*l*" loss and / or effective loss (that is delays, returns, damage of items in the chain etc.). With regard to losses and effective losses, it seems that, the vast majority of items in supply chains do not get lost (or in other words lost items are finally found). However, undamaged delivery within the promised time span is another issue and the delay and damage instances are more often than anticipated. To effectively model such losses one should include them in the supply chain as effective losses. This approach would disturb the simplicity of modelling but it cannot be helped otherwise. Further on, in this study, when losses are mentioned, effective losses would be implied and included as well. Embedding losses in the model is mainly necessary for the description of the middle mile, where most losses are expected and secondarily for the description of chain entry and the last mile phase.

Each interaction should have its own significance micro-factor in time units, depending on item's enthalpy and denoted as $\sigma_e(H)$, $\sigma_f(H)$, $\sigma_d(H)$ and $\sigma_l(H)$. In order to estimate if an interaction $i$ (= $e$, $f$, $d$ or $l$) is more important than others or even dominant over others, a dimensionless significance macro-factor $\Sigma_i(E)$ should be further introduced as

$$\Sigma_i(H) = \sigma_i(H) N(H) \tag{5}$$

where

$N(H)$ is the number of available interactors per unit time for interactions of type $i$ at enthalpy $H$.

Interactors may be described as receptors, when items are at the entry of the supply chain, mediators, when items during middle mile are forwarded to destination and finally couriers and when items are leaving the distribution post towards the client's address. A rather important type of interactors is the absorbers. It should be assumed that absorbers exist all along the supply chain and promote items loss and / or effective loss. Absorbers may be unnecessary procedures or bottle necks or other singular points in the chain (e.g. customs, transit, etc.). It is very important to mention that receptors, mediators and couriers could also be absorbers, when they malfunction or perform poorly in terms of items speed of transfer and items safety. The higher the significance macro-factor the most influencing the interaction in question. To determine an interaction rate, let us now define the flow of items into the supply chain as $w(H)$ in items per unit time. Then the interaction rate at energy $H$, $IR(H)$ (or interactions per unit time), could be written as:

$$IR(H) = \Sigma_i(H) w(H) \tag{6}$$

In addition, let us suppose that the supply chain is macroscopically in steady state and let us quantify the efficiency of the supply chain using a criticality ratio $k_{eff}$. This could be defined as

$$k_{eff} = \frac{\text{items delivered}}{\text{items entered}} \frac{1}{P_e \cdot p \cdot P_c} = k \frac{1}{P_e \cdot p \cdot P_{c_s}} = 1 \tag{7}$$

where

$P_e$ the losses escape probability during the fist mile entry phase

$p$ the losses escape probability during the middle mile forwarding phase, and

$P_c$ the losses escape probability during the last mile.

Finally, let us accept that for a supply chain there could be defined a core business area, where the density of forwarding mediators is big enough so that collection posts are in fact receptors and that last employed forwarding mediators themselves could serve as couriers emerging from a distribution post. Therefore, within the core area, collection posts and distribution posts are considered abundant and in sufficient density minimizing first and last mile effects. Definitely, the core business area has boundaries, while business itself extends beyond these boundaries. This calls for naming the boundary as the distribution front; naming the boundary as collection front does not seem correct since collection is a process initiated by an independent customer, while distribution is an obligation of the supply enterprise. Outside of the distribution front, all collection posts and distribution points and outposts would have a more discrete role and would be better described as first mile or last mile.

More necessary assumptions and definitions will be introduced in the course of this presentation.

## 3. First mile considerations

The first mile phase seems to be an hybrid part of the supply chain since, apart from productivity technicalities, it incorporates entrepreneurship and marketing parameters, really difficult to model. For example, one would not quantitatively explain customer's wishful thinking and expectations for submitting an item for shipment. Furthermore, the enthalpy property "$H_a$", as in Eq. 1, of the item to be collected is defined differently than the energy property "$H$", as in Eq. 4, of the item being shipped and delivered. In the case of "$H_a$", distance starts at the customer's origin address and ends at the collection post location. In the case of "$H$" distance starts at the collection post location and ends at the delivery address. Consequently, "$H_a$" and "$H$" are much different in nature and, for long chains, very different in terms of order of magnitude. Therefore, first mile should not be directly included in the proposed modelling. However, some discussion is necessary, since the first mile phase determines the items total flow $w$ into the supply system. Total flow at the interface just outside the collection post could be estimated as:

$$w = \int_0^{H_{a\,max}} w(H_a)dH_a \tag{8}$$

Conservation laws dictate that at the interface just inside the collection post, the same total flow could be estimated as:

$$w = \int_0^{H_{max}} w(H)dH \tag{9}$$

There is no apparent connection between function $w(H_a)$ and function $w(H)$. It should be mentioned that the actual value of total flow is irrelevant for this analysis since everything is considered in steady state and finally in terms of a criticality ratio and probabilities, as per Eq. 7, under the sole assumption that total flow is large enough for the supply chain to be somewhat profitable and, of course, small enough for the chain not to be overwhelmed. Therefore this approach eliminates any quantitative effects of flow. Total flow values obviously follow the normal distribution, inasmuch items present themselves at the collection post as Poisson events. Total flow is independent of enthalpy.

On the other hand $w(H_a)$ is of particular interest for the assessment of the performance and the peculiarities of a single collection post. Omitting the utopic homogeneous distribution with energy $H_a$, and ignoring all dystopic scenarios of multi peak distributions, the following hypotheses seem feasible.

(a) $w(H_a)$ follows the normal distribution. In other words its distribution is symmetric, i.e. mean $w(H_a)$ = median (or most probable) $w(H_a)$, thus, the collection post is well focused and tuned to items of particular energy, within its effective business activity radius. Items of different energies do appear for handling, however, and especially if the distribution has small standard deviation around the mean, they are a smaller part of the business. Seasonality and other causes of variation could be considered only as far as they influence standard deviation.

(b) $w(H_a)$ follows a non-symmetric distribution, much like the lognormal or the $\chi$ or the $\chi^2$ or the Weibull or even the Maxwell – Boltzmann distribution and its extensions; the latter have been introduced in econometrics with considerable success.

Since this negotiation is originating from reactor neutron physics, let us assume that the Maxwell-Boltzmann distribution is the one suitable for describing $w(H_a)$ of items that request shipment. It should be written

$$\frac{w(H_a)}{w} = \frac{2\pi\sqrt{H_a}}{(\pi T)^{3/2}} \exp\left(-\frac{H_a}{T}\right) \tag{10}$$

Equation 10 has certain advantages over other distributions. It presents a peak, which could be tuned to fit reality, it is biased towards lower energy as expected for the energy of items submitted for shipment and it can be adjusted to diverse markets or trends and / or changes of the collection post effective business radius. To this end, the "temperature" parameter $T$ (in enthalpy units) could be used to introduce market differentiations or to account for business model changes. For the purposes of this study, let us poetically call $T$ "market temperature". A plausible graph of this distribution is given in Fig. 1.

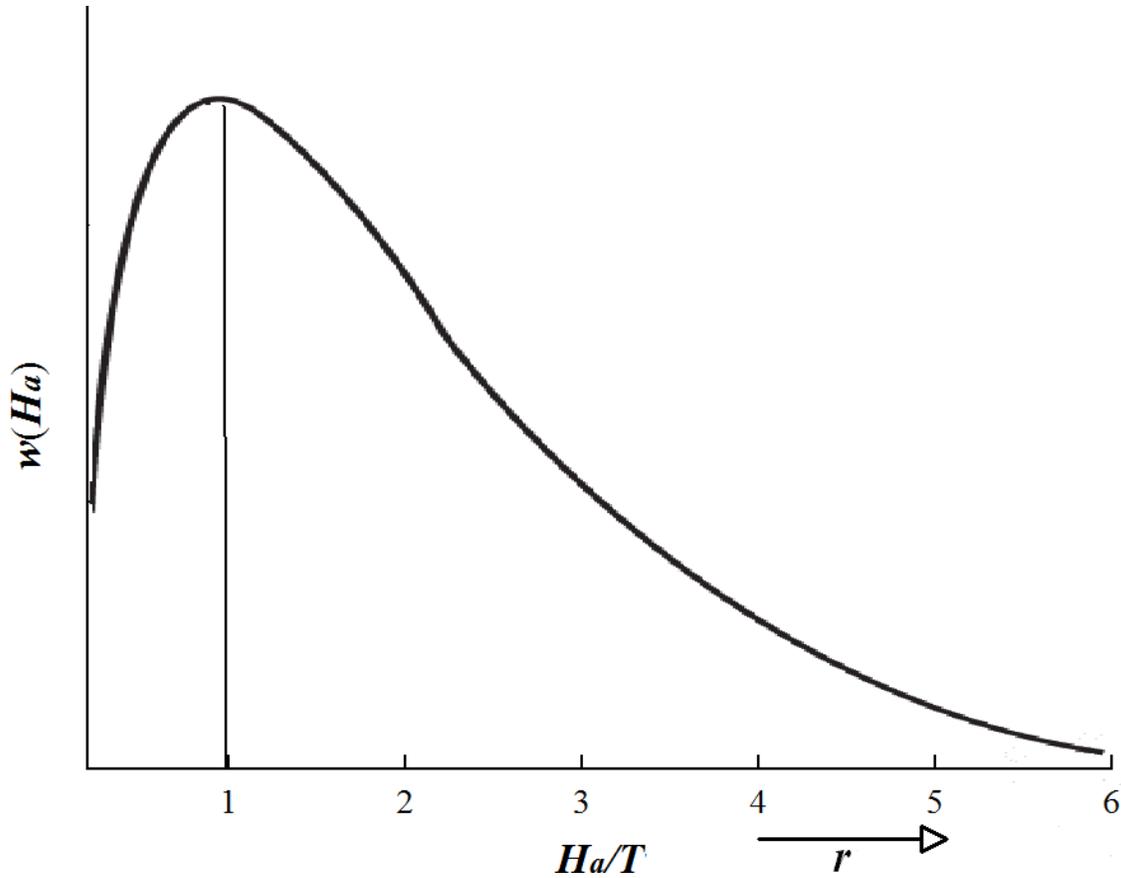

**Figure 1:** Plausible Maxwell - Boltzmann distribution of flow with enthalpy for items en route to collection; zero energy is defined at the collection post point; energies far greater than $T$ describe items originating at customers quite far away from the collection post.

## 4. Middle mile considerations

4.1 Losses escape probability $P_e$ during entry

It is necessary to review the performance of receptors at the entry of a supply chain so as to estimate initial losses. It has to be mentioned that entry to the supply chain is largely a single value enthalpy interaction taking place at $H = H_{max}$ and therefore concerning only total flow as given by Eq. 8, or as predicted by the plausible distribution at the first mile. Further, one cannot support that there would be no losses at entry since the service model at items receptors may vary either willingly or inadvertently. In addition, less suitable receptor types might accept items when the appropriate receptors are unavailable; in such a case if the suitable receptor is of type $i$ and has high $\sigma_{e,i}$ its available population $N_i$ would be minimum. Following these considerations it can be written that

$$P_e = \frac{w \sum_i \Sigma_{e,i}}{w \left( \sum_i \Sigma_{e,i} + \sum_j \Sigma_{l,j} \right)} = \frac{\sum_i \Sigma_{e,i}}{\left( \sum_i \Sigma_{e,i} + \sum_j \Sigma_{l,j} \right)} \qquad (11)$$

where

$i$ is the number of receptor types, and

$j$ is the number of absorber types

It is, once more, stressed that loss properties, should be attributed also to receptor interactors. Therefore $j$ should include erratically behaving receptors as well.

4.2 Forwarding mediators

The main purpose of forwarding is to stepwise drive items already in the supply chain down to the enthalpy level of the respective distribution post, i.e. down to the distance suitable for final delivery (last mile) processes. While items collection and items entry to the chain seem to be a matter of collection posts performance, effective forwarding is a matter of effective forwarding mediators. In this study a mediator is a term aiming to describe macroscopically complex processes such as item's transport, palletizing, warehousing and other logistics activities starting at the collection post and ending at the distribution post. In principle, the best step forwarding is connected to a mediator type, which could keep both processing time and logistic activities quantity and all associated costs to optimal minima. For example, if a supply process involves transportation through flights, it seems best for the respective mediator to employ the minimum possible number of flights and, cost plus flight frequency permitting, just a single flight.

At this point the need is raised to characterize the cooling down properties of a mediator type. To this end, one has to abandon enthalpy as the independent variable, since enthalpy spans within several orders of magnitude, making it useless to describe cooling down comprehensively. In reactor neutron slowing down theory, neutron lethargy $u_n$ is used instead of the neutron kinetic energy. Neutron lethargy is defined as:

$$u_n = \ln \frac{E_0}{E} \qquad (12)$$

where:

$E_0$ is the mean kinetic energy with which a neutron is born in a reactor, and

$E$ is the current neutron kinetic energy.

Neutron lethargy has many advantages: it is dimensionless; if $E_0$ is several orders of magnitude greater than $E$, the $u_n$ becomes almost independent of $E_0$; it describes the reduction from a very high $E_0$ to a very low $E$ far better than abstracting $E_0 - E$ or than dividing $E_0/E$.

Transferring neutron lethargy to the supply chain problem, supply lethargy $u_s$ could be introduced as:

$$u_s = \ln \frac{H_{max}}{H} \qquad (13)$$

Obviously, similarly to the neutron lethargy, supply lethargy is always positive, lethargy upon entry to the supply chain, at the collection post, is zero and lethargy upon exit (at the distribution post or at the client's address) is maximum. Supply lethargy could then be used to define, for a mediator type, the mean logarithmic reduction of item's enthalpy (or the mean increase of lethargy) per cool down step. Let $u_{s,out}$ be the lethargy after the cool down step and $u_{s,in}$ the lethargy before the cool down step. Then:

$$\xi_s = \overline{u_{s,out} - u_{s,in}} = \overline{\Delta u_s} = \overline{\ln \frac{H_{max}}{H_{out}} - \ln \frac{H_{max}}{H_{in}}} = \overline{\ln \frac{H_{in}}{H_{out}}} \qquad (14)$$

and

$$\xi_s = \overline{\Delta u_s} = \int_{H_{out}}^{H_{in}} \left( \ln \frac{H_{in}}{H'} \right) g(H_{in} \to H') dH' \qquad (15)$$

where

$H_{in}$ is the enthalpy, with which the item enters the cool down step,

$H_{out}$ is the enthalpy, with which the item exits the cool down step ($H_{out} <= H_{in}$),

$H'$ is any probable energy within [$H_{out}$, $H_{in}$] and, finally

$g(H_{in} \to H')$ is the elemental probability for the item in cooling down to reduce its enthalpy from $H_{in}$ to $H'$ using the investigated mediator type.

Unfortunately, while $g(H_{in} \to H')$ for reactor neutron slowing down is well understood and can be expressed in a deterministic manner, this does not seem to be the case for an item within a supply chain. Yet, an attempt is worth it because the analogies are really close. In Nuclear Engineering a mediator (called moderator) is usually a light nucleus available to scatter the neutron and thus reduce its kinetic energy. It has been derived that the minimum neutron energy that could be obtained after a scattering interaction is

$$E_{out} = \left( \frac{M-m}{M+m} \right)^2 E_{in} \qquad (16)$$

where

$M$ is the mass of the light nucleus, and

$m$ is the mass of the scattered neutron

It has also been derived that if the scattering nucleus is not too light, that is, if[1] $M > \zeta \cdot m$, $\zeta = 4$, $g(E_{in} \to E')$ is constant and equal to:

$$g(E_{in} \to E') dE' = \frac{1}{(1-a) E_{in}} dE' \qquad (17)$$

---

[1] Inequality factor $\zeta = 4$ is solely referenced for practical reasons. Scientists other than the author, might suggest that $\zeta$ should have a greater value.

where

$$a = \left(\frac{M-m}{M+m}\right)^2 \qquad (18)$$

Let us call, this constant elemental probability as an isotropic probability.

In order to transfer these conclusions to the supply chain processes, it is necessary to assign some kind of inertia *m* to the item that is subject to a forwarding interaction with a mediator. It seems that in this case *m* might be:

*m* = item's weight × item's volume × item's fragility × item's insurance value × ...  (19)

In the same manner, it is necessary to assign inertia (or better: inertia capacity) *M* to the mediator, in order to model an item's slowing down interaction similarly to the scatter interaction. Mediator's *M* is difficult to describe and quantify, nevertheless, instinct dictates that value of *M* should be greater and close to *m*. If the inertia capacity of the mediator is much greater than the inertia of the item in the chain, item will be ignored. The situation would compare to filing an individual's request into a complex bureaucratic organization. If the inertia capacity of the mediator is less than the inertia of the item, item will be handled the wrong way and, chances are that, its enthalpy will be increased. By application of these thoughts, the integral in Eq. 15 can now be calculated as:

$$\xi_s = 1 - \frac{(M-m)^2}{2M}\ln\left[\frac{M+m}{M-m}\right] \qquad (20)$$

Not to much surprise, $\xi_s$ of a single type of mediator is independent of item's enthalpy. If a set of mediator types available to serve concurrently and in parallel (e.g. warehousing plus palletizing) of different inertia capacities is used, as, most probably, this would be the actual situation, $\xi_{s\mu}$ of the mixture $\mu$ will be dependent on item's enthalpy as per the following short analysis:

Firstly, let $\Sigma_{f\mu}(H)$ as in

$$\Sigma_{f\mu}(H) = \sum_i \Sigma_{f,i}(H) \qquad (21)$$

be the total forwarding significance macro-factor for a set of mediator types. In this case, the total forwarding significance macro-factor could be accepted to be the sum of individual forwarding macro-factors.

Secondly, let $p_{f,i}(H)$ as in

$$p_{f,i}(H) = \frac{\Sigma_{f,i}(H)}{\Sigma_{f\mu}(H)} \tag{22}$$

be the probability of item with enthalpy $H$ to be forwarded with mediator $i$, given as a ratio of the forwarding significance macro-factor of $i$ to the total forwarding significance macro-factor.

In the case of a mediators mixture, Eq. 15 should be rewritten as:

$$\xi_{s\mu}(H) = \sum_i \int_{H_{out,i}}^{H_{in}} \left[\frac{\Sigma_{f,i}(H)}{\Sigma_{f\mu}(H)}\right]\left(\ln \frac{H_{in}}{H'}\right) g(H_{in} \to H') dH' \tag{23}$$

Consequently, mediators or mediators mixtures with greater $\xi_s$ could be considered as better than others. It is argued that this is not enough for the selection of a suitable mediator type. The single flight example is rather vivid: a single flight might transfer an item from start to finish with maximum $\xi_s$, however, if this flight operates once a fortnight, the item would be filed as lost much earlier. To account for such shortcomings and also for losses, the term mediation ability (or *MA*) could be defined as a product

$$MA(H) = \xi_{s\mu}(H) \frac{\Sigma_{f\mu}(H)}{\Sigma_{l\mu}(H)} \tag{24}$$

where

$$\Sigma_{l\mu}(H) = \sum_i \Sigma_{l,i}(H) \tag{25}$$

Mediation ability indicates that the best forwarding mediator or mediator mixtures at enthalpy $H$ are those with higher $\xi_{s\mu}(H)$, higher $\Sigma_{f\mu}(H)$ and lower $\Sigma_{l\mu}(H)$.

It is deduced that for optimal reduction of item's enthalpy within a supply chain, the chain should be thoroughly investigated and optimized for performance according to *MA(H)*. Furthermore, experience originating from reactor physics calls that optimal *MA(H)* selection has to be based on a product with high $\xi_{s\mu}(H)$. In other words: if one has to choose between mediators or mediator mixtures with the same $MA(H)$, the mediator with the greatest $\xi_{s\mu}(H)$ should be selected, since reduction of the item's enthalpy in large steps is highly desirable. Another not that obvious reason for such a selection relates to the fact that usually mediators with high $\xi_{s\mu}(H)$ are likely to have an non-homogeneous elemental (thus anisotropic) probability $g(H_{in} \rightarrow H')$ strongly biased to greater values for lower $H'$ energies. This particular anisotropy is also highly desirable, since items forwarding around their original enthalpy dictated by a homogeneous $g(H_{in} \rightarrow H')$ are being transported to destination in an orderly, however, relatively slower manner.

4.3 Losses escape probability $p$ during forwarding

In order to keep this part of the negotiation as simple as possible, the supply chain is supposed to employ just one mediator type. In addition and since there is the need to consider probabilities, several items subject to forwarding should be considered in order to establish the part, which successively continues till the distribution post or the final destination. To this end let $q(u_s)$ be the items per unit time available for cooling down at lethargy $u_s$. If these items are expected to be cooled down in an elemental manner and add $du_s$ to their lethargy, then, if cooling down is isotropic and done in several steps, the probability of such an elemental lethargy gain is $du_s/\xi_s$. Therefore,

$$q(u_s)\frac{du_s}{\xi_s}$$

would be the items per unit time, which would successfully increase their lethargy by $du_s$.

Given the definition of lethargy, as in Eq. 15, it is derived that

$$du_s = \frac{dH}{H} \tag{26}$$

Consequently

$$q(u_s)\frac{du_s}{\xi_s} = q(E)\frac{dE}{\xi_s E} \tag{27}$$

However, Eq. 25, indicates that, equivalently, a certain elemental quantity of items interactions with the mediator has taken place to produce this kind of elemental cooling down. This is given per unit time as:

$$\left[\Sigma_f(u_s)+\Sigma_l(u_s)\right]w(u_s)du_s = \left[\Sigma_f(H)+\Sigma_l(H)\right]w(H)\frac{dH}{H} \tag{28}$$

and it should be equal to expression 27; thus

$$\frac{q(H)}{\xi_s}dH = \left[\Sigma_f(H)+\Sigma_l(H)\right]w(H)dH \tag{29}$$

During this elemental slowing down process, there would be also an elemental loss $dq(H)$ for the items available for slowing down per unit time. Obviously

$$dq(u_s) = \Sigma_l(u_s)w(u_s)du_s \tag{30}$$

or

$$dq(H) = \Sigma_l(H)w(H)\frac{dH}{H} \tag{31}$$

Dividing Eq. 29 with Eq. 31, one gets

$$\frac{dq(H)}{q(H)} = \frac{\Sigma_l(H)}{\xi_s\left[\Sigma_f(H)+\Sigma_l(H)\right]}\frac{dH}{H} \tag{32}$$

or

$$d\left[\ln q(H)\right] = \frac{\Sigma_l(H)}{\xi_s\left[\Sigma_f(H)+\Sigma_l(H)\right]}\frac{dH}{H} \tag{33}$$

or

$$\ln\frac{q(H)}{q(H_0)} = \int_{H_0}^{H}\frac{\Sigma_l(H)}{\xi_s\left[\Sigma_f(H)+\Sigma_l(H)\right]}\frac{dH}{H}$$

(34)

The losses escape probability till enthalpy $H$ would be:

$$p(H) = \frac{q(H)}{q(H_{max})} \tag{35}$$

or

$$p(H) = \exp\left\{-\int_{H}^{H_{max}} \frac{\Sigma_l(H)}{\xi_s\left[\Sigma_f(H) + \Sigma_l(H)\right]} \frac{dH}{H}\right\} \tag{36}$$

If the aim is to estimate the loss escape probability $p$ of the forwarding process as a whole, from the collection post to the distribution post, then

$$p = \exp\left\{-\int_{H_c}^{H_{max}} \frac{\Sigma_l(H)}{\xi_s\left[\Sigma_f(H) + \Sigma_l(H)\right]} \frac{dH}{H}\right\} \tag{37}$$

However, if the aim is to pin point enthalpy loci, where there exist peaks in loss and delay probability [i.e. 1 - $p$ ($H$) probability] a gradual stepwise integration of Eq. 36 is required.

Of course, nuclear reactor physics go far deeper into an exhaustive description and calculation of $p$. There is no point to further transfer such analysis into the supply chain processes, since a lot of hypotheses have to be accepted, which might not be valid in logistics.

## 5. Last mile considerations

### 5.1 Delivery by diffusion

In the above discussion, it was silently accepted that both first mile and middle mile phases are dimensionless and shapeless in terms of space. As far as the first mile phase is concerned, distances from customer's address to the collection post certainly play a role, however, the distance from which an item could be collected in order to be accepted in the chain relates more to the business model and marketing of the supply interface. Therefore, the proposed modelling cannot take into account any dimensions or shape related to the first mile phase. Subsequently, it was assumed that the middle mile phase is irrelevant of geometry and dimensions. This, of course, is not quite true, however, the supply enterprise would be obliged to serve an already accepted item as promised thus implying that their dimensional capacity is large enough (thus, in particular, infinite) to accommodate their promises for this item. Any losses attributed to exits from the enterprises' business space balloon are recorded as just losses with no further description or justification. The proposed modelling considers the supply business space indirectly, within the variable of enthalpy. Within the Nuclear Engineering science slowing down distances and interaction volumes are considered explicitly and bare a physical meaning, which seems transferable to supply chains but this would complicate the introductory character of the so far presented arguments.

Within the last mile phase, shape and dimensions seem to play a greater role and should, in principle, incorporated in the modelling under the assumption that the items enthalpy has become too low and that enthalpy is distributed close to a peak. The items delivered in this energy situation might be considered as of single enthalpy value. Single value enthalpy finite items propagation in space is known to call for the application of a diffusion model. If instead of the flow $w$, the flux $\varphi$ in items per unit time per unit surface is considered, the formulation for a steady state, and independent of enthalpy, diffusion situation is well known and straightforward

$$D(\vec{r})\nabla^2\varphi(\vec{r}) - [\Sigma_a(\vec{r}) - \Sigma_l(\vec{r})]\varphi(\vec{r}) + s(\vec{r}) = 0 \tag{38}$$

where

$D(\vec{r})$ is the item diffusion constant over the enthalpy range of their distribution (in surface units)

$\Sigma_d(\vec{r})$ is the item delivery interaction macro-factor

$\Sigma_l(\vec{r})$ is the item losses interaction macro-factor, and

$s(\vec{r})$ is the item sources as distributed in the diffusion media

Solutions of Eq. 38 are well known and depend on the geometry of the flux source and the geometry of the diffusion space. However, in delivery practice, diffusion space could be defined as of too big or of infinite dimensions, in order to accommodate deliveries of excess enthalpy properties. According to the assumption of a core business area boundary and depending on the forwarding mediators employed within the core, items forwarded to similar geographically close addresses would arrive to closely placed but, anyway, different distribution posts forming the distribution front. Therefore, it would be qualitatively and even quantitatively better to assume that there is a distribution surface source at $x = 0$ (with $y \to \infty$, $z \to \infty$), which releases courier interactors to serve the last mile at distance $x$. Equation 38 becomes one-dimensional and source-less for $x > 0$

$$\frac{d^2\varphi(x)}{dx^2} - \frac{1}{L^2(x)}\varphi(x) = 0, \; x > 0 \tag{39}$$

where $L(x) = \sqrt{\dfrac{D(x)}{\Sigma_d(x) + \Sigma_l(x)}}$

would be the diffusion length of the one-dimensional delivery process.

The difficulty with Eq. 39 is clear: The media, within which delivery evolves, is not homogeneous or better, it is not served homogeneously, therefore all parameters in the equation would depend on distance from the source. The discussion could be promoted qualitatively, if it is accepted that $L(x) = L$, not dependent on $x$. Eq. 39 could then be rewritten as

$$\dfrac{d^2\varphi(x)}{dx^2} - \dfrac{1}{L^2}\varphi(x) = 0,\ x > 0 \qquad (40)$$

It's solution is known to be

$$\varphi(x) = A e^{-x/L} \qquad (41)$$

and its physical meaning is that, at distances greater than 5 or 6 times the diffusion length, the last mile delivery phase is not feasible any more. Given that the density of couriers interactors in great distances from the items source, let alone their performance, would rapidly deteriorate, it can be stated that in reality

$$\dfrac{L^2(x)}{L^2} < 1,\ \lim_{x \to \infty}\dfrac{L^2(x)}{L^2} \to 0 \qquad (42)$$

thus reducing significantly the distance from the distribution surface, within which last mile processes can be accomplished. This is obvious for big supply enterprises operating for profit. Usually, their collections and more significantly, their deliveries are processed really close or within the core business area, unless customers are willing to pay surcharges. Subsidized supply chains, similar to state postal services do deliver far beyond their core area boundary at a time cost.

One would argue that, nevertheless, distribution points or distribution outposts do exist in a delivery system and should be modelled explicitly. To address this, a couple of approaches seem feasible: (a) Equations 41 and 42 are physically true even for a point source of items for delivery, provided that the distance from the point is large enough, and (b) at the bottom line, these points or posts could be considered as courier types. A third approach could be to consider distribution points more closely, with effectiveness within a 2D circular space of radius $r \to \infty$. The source of items to be delivered in the last mile would then be point like. The solution of the diffusion equation would though involve the zero order Bessel function of the first kind, which, unless border $r$ is finite, would yield results difficult to connect with logistics applications. The third approach does exist in nuclear reactor physics for finite $r$ and is used for the calculation of the so-called coupling of different neighboring reactor cores.

5.2 Enthalpy distribution

Similarly to the first mile phase an enthalpy distribution could be attributed to $w$ in the last mile phase. This, once more, might be of particular interest for the assessment of the performance and the peculiarities of a distribution front. Omitting again the homogeneous distribution with enthalpy $H_c$, and ignoring unlikely multi peak distributions, the following hypotheses could be discussed.

(a) $w(H_c)$ follows the normal distribution, thus, the distribution is well focused, well balanced and tuned to items of particular energy, within an effective business activity radius. However, one should abstain from the normal distribution hypothesis as far as arrival relatively close but not in the exact neighborhood of delivery addresses would, as previously mentioned, arise from a front rather than a point.

(b) $w(H_c)$ follows a non-symmetric distribution. The Maxwell – Boltzmann distribution would, without solid proof, be once more proposed for this, following the path that this negotiation is originating from reactor neutron physics. To no surprise this hypothesis bares particular weight soon to be explained. Caution should be paid that this distribution hypothesis, although mathematically equivalent with that of the first mile, its physical meaning is quite different. If the Maxwell-Boltzmann distribution is the one suitable for describing $w(H_c)$ of items that request delivery, then it should be written

$$\frac{w(H_c)}{w} = \frac{2\pi\sqrt{H_c}}{(\pi T)^{3/2}} \exp\left(-\frac{H_c}{T}\right) \tag{43}$$

This distribution has two main differences from that of Eq. 10 of the first mile phase: (i) It should incorporate losses during deliveries, an issue which seems irrelevant in the first mile phase, especially if the supply enterprise follows a drop-off model for collecting items, (ii) sources of items for delivery have a given geometry (i.e. the distribution front), while items sources for collection are well distributed in space, thus effect of space could be ignored in the first mile phase. These two essential differences would have an impact on the market temperature $T$ of the proposed distribution $w(E_c)$. Let us first consider the effect of losses, which are more likely to occur at lower energies (i.e. at greater distances from the distribution front). This would lower the distribution peak and probably push it to higher temperatures, or equivalently to higher enthalpy, where the supply enterprise has more control over the delivery processes. Let us now consider the effect of sources. The sources occur at high enthalpy thus enhancing the distribution peak and probably attracting it, again, to higher temperatures. According to nuclear reactor theory, the supply enterprise could have control over this phenomena, in quite a simple manner, which could be described heuristically by a temperature multiplier factor $\theta$ defined approximately as

$$\theta = 1 + \frac{\Sigma_d(T) + \Sigma_l(T)}{\xi_s \Sigma_f(H_{b\min})} \tag{44}$$

where

$\xi_s \Sigma_f(H_{b\min})$ could represent the performance properties of the last forwarding mediator.

The high enthalpy tail of Eq. 43 indicates items that successfully reached a distribution front and are subject for delivery, however, it would have been far better if the forwarding phase could place them to a front closer to their final destination. It is deduced from this last remark and also from the proposed distribution in Eq. 43 and commentary (i) and (ii) leading to the temperature multiplier factor in Eq. 44, that the supply enterprise, can, certainly at a cost and according to their abilities and capabilities, exercise enough control on the lower values of enthalpy $H_b$ (or equivalently on the start values of enthalpy $H_c$). In other words, the supply chain can adjust the distribution front's distance from the clients addresses to their will, thus making the boundary between forwarding enthalpy and last mile enthalpy not as clear, as already mentioned in section 2. A plausible look of the distribution as suggested in Eq. 43 is given in Fig. 2.

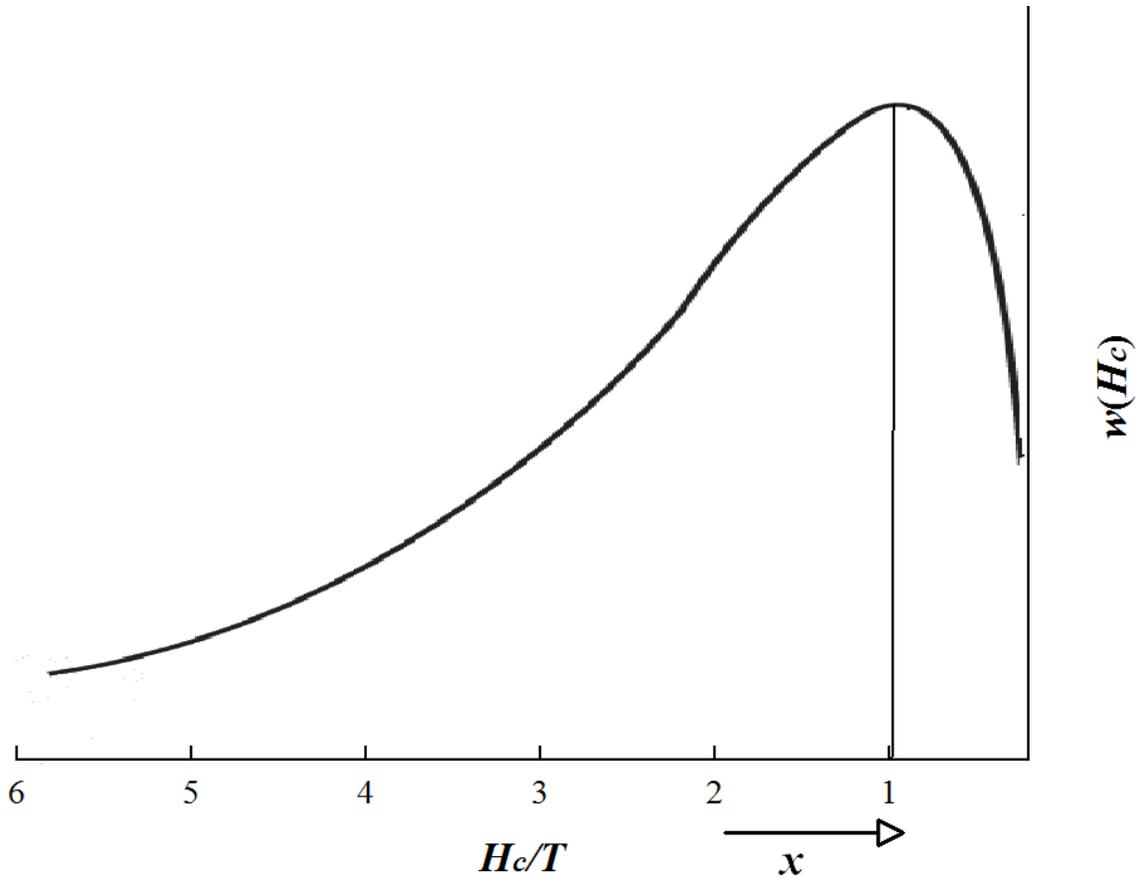

**Figure 2:** Plausible Maxwell - Boltzmann distribution of flow with energy for items in the last mile; zero energy is defined at the delivery address; energies far greater than $T$ describe items originating at a distribution front quite far away from the clients address.

5.3 Losses escape probability $P_c$ during delivery

It is necessary to review the performance of courier interactors at the exit (or at the end) of a supply chain so as to estimate final losses. It has to be mentioned that service at the last mile phase is enthalpy dependent and cannot be considered on the bases of total flow as in the entry point; the distribution with enthalpy should be employed. Further, one cannot support that there would be no losses at exit since the service model with items couriers may vary either willingly or inadvertently. In addition less suitable courier types might accept items when the appropriate couriers are unavailable. Finally, courier service performance seems to depend on distance from the distribution front as far as their density within the delivery space is getting less with distance. Following these considerations it can be written that losses escape probability $P_c$ during delivery is

$$P_e = \int_0^{H_{b\max}} \frac{\sum_i \Sigma_{d,i}(H_c) w(H_c) dH_c}{\left[\sum_i \Sigma_{d,i}(H_c) + \sum_j \Sigma_{l,j}(H_c)\right] w(H_c) dH_c} \quad (45)$$

where

$i$ is the number of courier types, and

$j$ is the number of absorber types

It is, once more, stressed that loss properties, should be attributed also to courier interactors. Therefore $j$ should include erratically behaving couriers as well.

## 6. Discussion and outlook

The previously described model is proposed "as is" and its connection to the reality of supply chains remains to be proven. Units within the model have been kept essentially limited, however proofing processes might require different arrangements. Proof, or the simpler proof of concept or even demonstration examples are difficult to organize and process. Main reason is that, for this purpose, there is the need of data, or more accurately big data, which are of proprietary sources closely related to profitable business models and thus difficult to share publicly. An initial investigation could focus on publicly available pricing details of courier enterprises operating in wide business spaces. This could allow for an initial determination of the core business areas; these, given some adjustments depending on the data processing method, should agree with what is largely known both in Europe and the U.S. Core areas boundaries would define the distribution front. Space outside core areas should extend to infinity, however as indicated by Eq. 41 and 42 not much could be done to serve it sufficiently and further from an extension length, beyond which all diffusion and most probably many convection equations[2] seize to apply. Further to this investigation, research could be promoted by a comparison analysis of the pricing lists regarding regular and premium deliveries. The extra cost would allow to estimate losses escape probabilities, both in the core area and also in the area beyond the distribution fronts. Surcharges beyond the usual rate for destinations of similar distances from the core would assist in the verification of the distribution front.

---

[2] convection process: in this case boosting last mile to delivery by extra means at the cost of the courier

Big data themselves, if available, could initially be used for the investigation of the applicability of the Maxwell-Boltzmann distribution both at the first mile and also at the last mile phases. However, this would not be done efficiently, unless some decision is taken with regard to the exact definition of items enthalpy and market temperature. To this end, it seems that the cryptocurrency technology should prove useful and that an (internal to the supply) cryptocurrency value could be used instead of enthalpy.

Accepting an unknown as yet, cryptocurrency - enthalpy relation, certain modelling problems not specifically mentioned could be better addressed. For example, it would be answered, which may be the enthalpy of an item that, within the forwarding process, looses its identity and becomes a part of a larger package, or the enthalpy of an item that is "born" following the deconvolution of a larger package. Further, there would be defined the enthalpy at the distribution front, where items are more likely to be once again individually identified. The same or other internal cryptocurrency could address the problem of assigning inertia to served items and inertia capacities to forwarding mediators.

Having addressed the model unknowns of enthalpy and inertia, sets of systems of continuity equations may be formulated for several energies and solved for $\Sigma_i(H)$ estimations. In a subsequent phase $\Sigma_i(H)$ could be analyzed as products $\sigma_i(H)N(H)$, where, as already mentioned, $N(H)$ is the number of available interactors per unit time for interactions of type $i$ at enthalpy $H$. Solutions are expected to reveal if the total significance macro-factor for interactors operating concurrently or in parallel is actually a sum of individual macro-factors for single interactors or something different. The sum has been suggested silently many times in the proposed modelling, see, for example, Eq. 11, 21 - 25 and so on. According to the reactor physics experience, solutions are further expected to reveal dependence of the macro-factors on the inertia and/or the inertia capacity. Following this phase a deeper look into the $\Sigma_l(H)$, would allow for a careful accounting of the losses components as a sum of actual losses, delays and returns. Other sets of systems of continuity equations may be formulated for several distances outside the distribution front and solved for $L(x)$ estimations for last mile deliveries, bearing in mind that $x$, according to Eq. 41 and 42, should be always less than $(5\div6)\cdot L(x)$. Last but not least, losses escape probabilities would be analyzed as functions of enthalpy and / or cryptocurrency value.

The whole process would result to information sets, which could then be used for optimization processes. Several hints for optimization tools have been included in the previous sections. It has been, indirectly, suggested that the middle mile phase, seems to be the one, where the optimization margin might be greater. Special review seems necessary for decisions regarding forwarding mediators: which types, how many, what succession order, which succession points etc. This optimization, however, would be a result over an existing situation, which according to the proposed model, is analyzed in a deterministic manner. Yet, stochastic optimization could be based on the calculated results for the macro-factors thus allowing for the supply chain to improve itself by testing theoretically plausible scenarios.